\documentclass{aa}

\usepackage[]{epsfig}

\def\ltsima{$\; \buildrel < \over \sim \;$}
\def\simlt{\lower.5ex\hbox{\ltsima}}
\def\gtsima{$\; \buildrel > \over \sim \;$}
\def\simgt{\lower.5ex\hbox{\gtsima}}
\def\cgs{{erg cm$^{-2}$ s$^{-1}$}}
\def\ergs{{erg s$^{-1}$}}
\def\cm2{{cm$^{-2}$}}

\def\xnu{{$\chi^{2}(\nu)$}}

\def\xnn{{$\chi^{2}_{\rm \nu}$}}
\def\fhx{{$F_{2-10}$}}
\def\lum{{$L_{2-10}$}}
\def\p1{{Paper I}}

\def\xmm{{\em XMM--Newton}}
\def\chandra{{\em Chandra}}

\def\nhgal{{N$_{\rm H}^{\rm Gal}$}}
\def\nh{{N$_{\rm H}$}}
\def\nhz{{N$_{\rm H}^{\rm z}$}}
\def\chandra{{\em Chandra}}

\def\xmm{{\em XMM--Newton}}
\def\nhgal{{N$_{\rm H}^{\rm Gal}$}}
\def\nh{{N$_{\rm H}$}}  
\def\epic{{\em EPIC}}
\def\mosuno{{\em MOS1}}
\def\mosdue{{\em MOS2}}
\def\pn{{\em PN}}
\def\mos{{\em MOS}}

\def\f14{{10$^{-14}$}}
\def\f13{{10$^{-13}$}}
\def\rx{{RX~J1343.4$+$0001}} 
\begin{document}
 
\title{The XMM-Newton spectrum of the high-z optically-obscured\\ QSO RX
J1343.4+0001:  a classic radio quiet QSO\thanks{Based on observations
with  XMM-Newton, an ESA Science Mission with instruments and
contributions  directly funded by ESA Member states and the USA
(NASA)}}
 
\author{E.~Piconcelli\inst{1}, M.~Guainazzi\inst{1}, M.~Cappi\inst{2}, E.~Jimenez--Bailon\inst{1}, N.~Schartel\inst{1}}

\titlerunning{The XMM-Newton spectrum of the high--z optically--obscured QSO
RX~J1343.4+0001} \authorrunning{E.~Piconcelli et al.}
 
\offprints{epiconce@xmm.vilspa.esa.es}
 
\institute{XMM-Newton Science Operations Centre (ESAC), Apartado 50727, E--28080 Madrid, Spain \and IASF/CNR, via Piero Gobetti 101, I--40129 Bologna, Italy}
 
\date{}
 
\abstract{We present a 30 ks \xmm~observation of the $z$ = 2.35 Type II
radio quiet quasar \rx. These data provide the first good quality X--ray spectrum 
for this object. 
We measured a continuum slope $\Gamma$ = 1.85$\pm$0.10 with only an upper limit
on the column density of the absorbing material of \nhz~\simlt 10$^{22}$ \cm2~as well as an Fe K$\alpha$
emission line at the 3$\sigma$ confidence level.
We do not find therefore a highly absorbed object nor a truly flat spectrum as suggested  on the
basis of previous less sensitive {\em ROSAT} and {\em ASCA}
measurements. The \nhz~upper limit is fully consistent with the optical extinction
3$<A_V <$10 inferred from IR observations. The Fe K$\alpha$ line is consistent with 
fluorescence from neutral iron  and, noteworthy, is one of the most distant observed so far.
The X--ray spectral properties of \rx~agree well with the steep continuum slope ($\Gamma \approx$ 1.9) 
being independent of increasing redshift and luminosity as inferred by  X--ray studies of large samples of RQ QSOs.
 }
 
\maketitle
 
\section{Introduction}
In the framework of unified models, the differences between Type I and
Type II Seyferts are explained by the existence of an  obscuring
screen (the so-called ``torus''), which prevents us from observing the
inner regions around the black hole  (i.e. the accretion disk and the
broad line region), if our line of sight intersects it (Antonucci
1993).  This picture (which is thought to hold also for QSOs) 
implies that Type I and II AGNs are essentially the same objects
viewed at different inclination angles. Nonetheless, the existence of
high--luminosity analogues of Seyfert IIs (i.e. with only narrow
emission lines in their optical spectra) has been hotly debated during
the last decade, and only with the advent of new generation X-ray
telescopes \xmm~ and \chandra~has a handful of these objects  been
unambiguously detected in the optical follow--up of deep X--ray
surveys (e.g. Norman et al. 2002; Padovani et al. 2004; Fabian et al. 2003 for a review).

Most of the studies carried out so far on the X-ray spectral
properties of QSOs  concern optically-selected (i.e. not obscured)
samples at $z$ \simlt~1 (see e.g. Zamorani et al. 1981; Williams et
al. 1992; Fiore et al. 1998).
At higher redshifts (i.e. $z \gg$ 1) the situation is more uncertain
because of the limited number of sources accounted for so far.  Moreover,
since at a given optical luminosity  ``radio quiet'' (RQ) QSOs are
$\sim$ 3 times less X-ray luminous than  ``radio loud'' (RL) QSOs, most of
the spectroscopically analyzed QSOs at higher redshifts were RL.  So
the average spectral properties observed for  RL QSOs at low $z$
($\langle\Gamma\rangle_{RL} \sim$ 1.6 and intrinsic \nh~exceeding the
Galactic value) were confirmed  to persist up to $z \sim$ 3-4  by
{\it ASCA} and {\it ROSAT} observations  (Cappi et al. 1997; Reeves \&
Turner 2000, hereafter RT00).

On the contrary, spectral properties of high-$z$ RQ QSOs were poorly
constrained before the launch of \xmm~and~\chandra. Early {\it ASCA}
results (Vignali et al. 1999), which suggested a flattening of the
average slope with increasing redshifts, have been recently revised after
the results of individual spectral analysis of QSOs at $z \sim$ 2
detected in a shallow \xmm~survey (Piconcelli et al. 2003, hereafter P03) and
stacked-spectra analysis of very high-$z$ (i.e. z $\approx$ 5--6)
objects (Vignali et al. 2003).  
These findings indicate that no change in the X--ray continuum slope with redshift occurs.
Such a  trend is also confirmed by recent optical surveys (Fan et al. 2004) which find  no significant evolution in the
average emission line and continuum properties of $z \sim$ 6 QSOs compared to those of lower redshift
samples of QSOs.\\

It is worth stressing  that these results are based on
samples of broad line QSOs.  The X-ray spectral properties of type II
QSOs, however, remain almost unexplored owing to their very faint
X-ray fluxes.  Furthermore the existence of such a large population of
high-luminosity ($L_X >$ 10$^{44}$ \cgs) obscured objects is also
required in all the synthesis models of the  cosmic X-ray background
in order to reproduce the source counts in the 2--10 keV band (Gilli
et al. 2001).

In this paper we report on the \xmm~observation of RX J1343.4$+$0001
($z$ = 2.35; Almaini et al. 1995, hereafter A95).  
Its relative brightness ($F_{0.5-10}
\approx$ 2 $\times$ 10$^{-13}$ \cgs) makes \rx~an excellent candidate to get better insight in the X--ray spectral properties of high-$z$ RQ QSOs.

\section{RX J1343.4$+$0001}
\label{sec:rx}
\rx~was
discovered during the optical follow-up of a deep {\it ROSAT} survey
and classified as a Type II QSO  because of its high luminosity ($L_X \sim$ 2
$\times$ 10$^{45}$ \ergs) and the lack of any broad lines (i.e. with
$v >$ 900 km/s) in its optical spectrum (A95). The {\it ROSAT} 0.1-2
keV low-statistics spectrum was fitted by A95 with an unabsorbed power
law model with $\Gamma \sim$ 1.5.

Georgantopoulos et al. (1999, hereafter G99) presented the results of
near-IR and {\it ASCA} observations of this QSO. They
detected a broad H$\alpha$ emission line redshifted at 2.2 $\mu$m:
therefore this object is not a pure QSO 2, but, more properly, a
``Type 1.9'' QSO, i.e.   a luminous analogue of a Seyfert 1.9 galaxy,
considering the source--frame optical spectra.  The {\it ASCA}
spectrum is hard and either a flat ($\Gamma$ $\sim$ 1.3)  or a
steeper ($\Gamma$ $\sim$ 1.9) absorbed (\nh~$\sim$ 10$^{23}$ \cm2)
power-law model  yielded acceptable fits.  The limited photon
statistics prevented a firm conclusion on the
spectral shape of this source. 
Similar conclusions were reached by G99 who also added {\it ROSAT} data in their analysis.
\rx~was also included in the {\it ASCA} QSOs survey presented by
RT00. These authors reported the
detection (at 90\% confidence level) of a Fe K$\alpha$ line at 6.4 keV
with a rest--frame equivalent width EW = 432$_{-358}^{+435}$ eV.

\section{XMM--Newton observation and data reduction}
\rx~was observed by \xmm~(Jansen et al. 2001 and references therein)
on January 29, 2004 for 35.4 ks. The \epic~\pn~(Turner et al. 2001) and \mos~(Struder et al. 2001) observations
were carried out in the full frame mode using the Thin filter.
\xmm~data were processed with SAS v6.0.  We used the  EPCHAIN
and  EMCHAIN tasks for processing the raw \pn~and \mos~data files,
respectively, in order to generate the relative linearized event
files.  X--ray events corresponding to patterns 0--12(0--4)  for the
\mos(\pn)~cameras were selected. We employed the most updated
calibration files at the time the reduction was performed (June
2004). Flickering and bad pixels were removed.  The event
lists were furthermore filtered to ignore periods of high  background
flaring  according to the method presented in Piconcelli et
al. (2004a) based on the cumulative distribution function of
background lightcurve count-rates.  After this data cleaning, we
obtained as net exposure time 24.9, 27.8, 32.4 ks for \pn, 
\mosuno~and\mosdue, respectively.  Given the current calibration
uncertainties,  we discarded  \pn(\mos) events below 0.3(0.6) keV and
above 12(10) keV.  \epic~spectra were binned to a minimum of 20
counts/bin to facilitate the use of the $\chi^{2}$ minimization
technique in the spectral fitting.  All fits were performed using the
XSPEC package (v11.3). 
%In Fig.~\ref{f1} the \pn~broad--band
%background--subtracted binned spectrum is shown.
\pn~and \mos~data
were then fitted simultaneously.  The quoted errors in the model
parameters correspond to a 90\% confidence level for one interesting
parameter ($\Delta\chi^2$ = 2.71; Avni 1976). All luminosities are
calculated assuming a $\Lambda$CDM cosmology with ($\Omega_{\rm
M}$,$\Omega_{\rm \Lambda}$) = (0.3,0.7) and a Hubble constant of 70 km
s$^{-1}$ Mpc$^{-1}$ (Bennett et al. 2003).

\section{The X--ray spectrum}\label{spectrum}
%%%%%%%%%%%%%%%%%%%%%%%%%%%%%%%%%%%%%%%%%%%%%%%%%%%%%%%%%%%%%%%%%%%%%%%%%%%%%%%%%%%%%%%%%%%%%%%%%%%%%%%%%%%%%%%%%%%%%%
We initially fitted the \epic~spectrum with a power law model modified
by  Galactic absorption (\nhgal~=2.6 $\times$ 10$^{20}$ \cm2;
A95). This fit (PL hereafter) turned out a photon index  $\Gamma$ =
1.82$^{+0.10}_{-0.09}$ with an associated \xnu~=41(47).  

\begin{figure}\centering
\epsfig{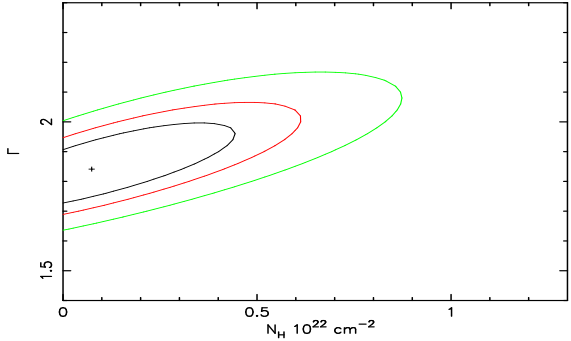}
   \caption{Confidence contour plot showing the QSO photon index ($\Gamma$) against the
intrinsic rest--frame column density (\nhz). The contours are at 68\%, 90\% and 99\% confidence levels, respectively, for two interesting parameters. The column density of the Galactic foreground absorption is \nhgal~= 2.6 $\times$ 10$^{20}$ \cm2. }
   \label{f2} 
\end{figure}

In order to test the presence of heavy intrinsic obscuration in \rx~as claimed
by previous X--ray studies, 
we added an additional absorption component due to neutral gas to the PL model.  No intrinsic absorption was significantly detected and 
the upper limit on the column density in the QSO rest--frame ($z$ = 2.35) was
\nhz $<$  5 $\times$ 10$^{21}$ \cm2.  In Fig.~\ref{f2}  the
two--dimensional contour plot in the parameter space \nhz--$\Gamma$ is shown. 
Our \epic~data place tighter than ever limits on the slope of the primary continuum as well as on the 
amount of the absorption in \rx. 
Note that assuming the RT00 best fit model (i.e. $\Gamma$ = 1.9$\pm$0.4 and \nhz~4.8$^{+6.5}_{-4.1}$ $\times$ 10$^{22}$ \cm2) yields \xnn~$\approx$ 4.2.
We also tested the possibility that the obscuration occurs in an `warm' (i.e. ionized) material.
Even if the X--ray opacity of such a gas is lower than neutral, a warm absorber with internal dust could, however, 
significantly obscure the optical/UV band (as suggested
in Komossa \& Fink (1997) for the Seyfert 1.8 NGC3786).
This fit, performed using the {\tt ABSORI} model in XSPEC,  yielded no statistical improvement with respect to
the model with cold absorption. The resulting column density of the warm gas was
\nh~$<$ 1 $\times$ 10$^{22}$ \cm2, while its ionization parameter remained unconstrained with a best-fit value 
$\xi \sim$0 \cgs. 
 
%A simple power law parameterization appear  s therefore to provide an
%adequate description of the $\sim$ 1--40 keV (rest--frame) continuum
%emission of \rx~which does not show the presence of any obvious
%curvature both in the soft and hard band.

\begin{figure}\centering
\epsfig{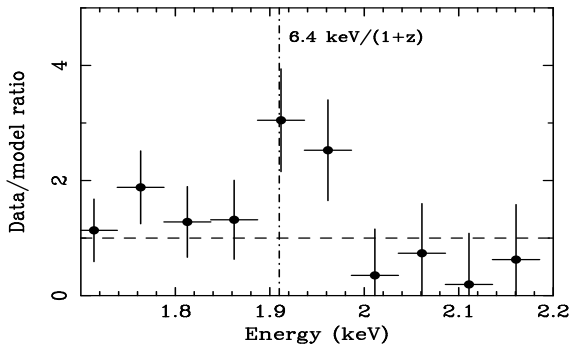}
\caption{A close--up of the ratios of the \pn~spectral data to a simple power law model. 
The plot shows the excess at $\sim$ 1.9 keV corresponding to $\sim$ 6.4 keV in the QSO rest--frame.}
\label{fig:line} 
\end{figure}

Visual inspection of the data--to--model ratio residuals (see
Fig.~\ref{fig:line}) suggests the presence of an emission feature at $\sim$ 2
keV (i.e. $\sim$ 6.4 keV in the QSO rest--frame).  This spectral
feature was already marginally detected by {\it ASCA} (RT00) and it is
associated to fluorescence emission from the K--shell of iron.  Thus
we introduced a narrow Gaussian line in the fitting model to account
for it.  The fit  gave a rest--frame energy of the line at E =
6.42$^{+0.08}_{-0.21}$ keV (EW = 460$^{+330}_{-290}$ eV in the QSO--frame) with a statistical
improvement significant at 98.5\% confidence level. The centroid of
the line corresponds to low ionization states, i.e. Fe I--XVIII
(Makishima 1986). The resulting slope of the underlying continuum was $\Gamma$ = 1.85$\pm$0.10.
Furthermore we checked (and ruled out) the possibility that background subtraction affects the detection and/or the strength of this line
by a visual inspection of the source and background superposed spectra.
We also fitted the line leaving the
$\sigma_{K\alpha}$ parameter free to vary, which yielded a value of
$\sigma_{K\alpha}$ $\sim$ 0 (with an upper limit of
$\sigma_{K\alpha}<$ 1.3 keV) and no statistical improvement in the
goodness of fit. The velocity width of the line is not well
constrained  because of  the limited number of photons collected
from this faint QSO. Further investigations on the possible ``broadness'' of this line 
require better data quality.\\

We measured a 2--10 (0.5--2) keV observed flux of
6.5$\pm$0.1(4.0$\pm$0.2) $\times$ 10$^{-14}$ \cgs, which corresponds to a luminosity
of $\sim$ 2.2 (1.1) $\times$ 10$^{45}$ \ergs~once corrected for
Galactic absorption.  During this \xmm~observation the 0.5--10 keV flux
of \rx~was therefore a factor of 40\% lower than measured
by {\it ASCA} on 1996. Assuming that the black hole produces X--rays with an efficiency of 0.01 $\times$ $L_{\rm Edd}$ 
(Norman et al. 2002) we can estimate a  M$_{BH} \approx$ 3 $\times$ 10$^{9}$ M$_\odot$ which lies in the typical range
inferred for high--$z$ QSOs (Dietrich \& Hamann 2004).

\begin{figure}\centering
\epsfig{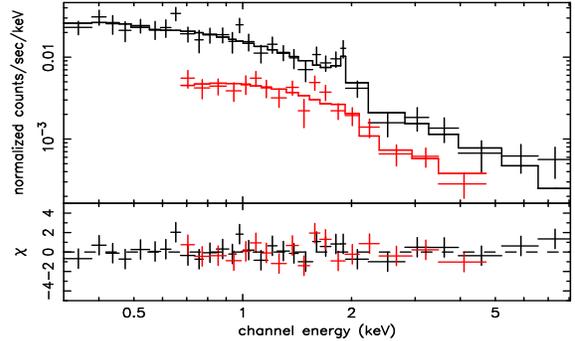}
  \caption{\epic~\pn~and \mos~spectra  of \rx~when the best fit model (i.e. power law $+$ Gaussian line) is applied.
 The lower panel shows the deviations of the observed data from the model.}
   \label{f4} 
\end{figure}

\section{Discussion}
\subsection{X--ray continuum and absorption}

The \epic~observation presented here provides the first good--quality
X--ray spectrum of this ``Type 1.9 QSO'' as  it allows us to
strongly constrain both the slope of the primary continuum and the
column density of the absorbing material,   $\Gamma$ = 1.85$\pm$0.10 and \nhz~\simlt 10$^{22}$
\cm2, respectively. Thanks to the
unprecedented large collecting area of \xmm, it is therefore possible
to rule out both  hypotheses about the X--ray nature of
\rx~suggested on the basis of low--sensitivity {\it
ROSAT} and {\it ASCA} measurements, i.e. a truly flat or a highly
obscured spectrum (e.g. Sect.~\ref{sec:rx}).   Note that Georgantopoulos et al. (2004) derived similar values for
these spectral parameters (i.e $\Gamma$ = 1.77 $\pm$ 0.36 and \nhz~$<$ 6 $\times$ 10$^{22}$ \cm2)
from an off-axis shallow \epic~exposure of \rx. We also performed a hardness ratio (HR) analysis of a 10 ks \chandra~observation of this QSO.
The HR for this source is slightly lower than the corresponding one calculated using $\Gamma$ = 1.9 and \nhz~= 10$^{22}$ \cm2.
This finding implies that a column density value larger than  10$^{22}$ \cm2 is  also ruled out on the basis of these \chandra~data,
taken when the source was at a similar flux level as the present \epic~observation.\\ 

Previous {\it ASCA} claims (G99; RT00) for  a large
amount of \nhz~($\sim$ 10$^{23}$ \cm2) appear to be due to the
limited sensitivity and the narrower energy range affecting these
measurements. We rule out a possible contamination due to another source present in the
ASCA errorbox: no additional X--ray source lies within the 1 arcmin 
box centered on the QSO. 

Some recent works based on \xmm~observations (Ferrero \& Brinkmann 2003;
Grupe et al. 2004) also do not confirm earlier {\it ASCA} results about the
presence of a strong  absorption toward high--$z$ QSOs.
Nevertheless, even though affected by large errors, the results found by A95 (i.e. a
95\% upper limit of $\Gamma <$1.98 and  a value of  \nhz~$<$ 10$^{22}$
\cm2) for the {\it ROSAT} spectrum are consistent
with ours.
Furthermore, the value obtained for the photon index is 
similar to the canonical one measured for RQ QSOs at 0.01\simlt~$z$\simlt~2 (e.g. RT00).  This
result is in agreement with the observational evidence that the
mean X--ray spectral shape of QSOs does not show any variation over
$z$ (P03; Vignali et al. 2003).

The upper limit on the column density (\nhz~\simlt~ 10$^{22}$ \cm2) is
fully consistent with the optical extinction.   On the basis of the
results from an IR observation (i.e. $B-K$ = 5.4 and presence(absence)
of the broad H$\alpha$(H$\beta$) line), G99 concluded  that this QSO
is obscured by a moderate amount of dust, with a lower limit of the
photoelectric extinction of $A_V >$3 (excluding, however, a reddening
much higher than this value i.e. $A_V <$ 10 as suggested by
Georgantopolous et al. 2003).  Such a limit on the reddening to the
BLR implies a range for the column density of 5 $\times$ 10$^{21}$
\simlt~\nhz~\simlt~2 $\times$ 10$^{22}$ \cm2~according to the formula
\nh/$A_V$ = 1.79 $\times$ 10$^{21}$ \cm2~mag$^{-1}$ (Predehl \&
Schmitt 1995).   On the one hand, this suggests that X--ray and
optical/UV obscuration  likely occur in the same matter
(e.g. the putative torus invoked in the AGN Unified models); on the
other hand, it rules out the hypothesis of  a gas-to-dust ratio along the
line of sight several times larger than that of the Milky Way (G99).
\rx~therefore poses a notable exception among intermediate type
1.8-1.9 AGNs which usually have a low $A_V$/\nh~as pointed out by
Maiolino et al. (2001) and Granato et al. (1997). Such a possibility
has also been proposed to explain the existence  of X-ray absorbed
broad line QSOs (Akyiama et al. 2000).  However, recent
\xmm~results (P03; Akylas et al. 2004) do not yield a large
population of these objects and, moreover, there is growing evidence
that obscuration in  broad line QSOs is basically due to ionized
(instead of ``cold'') material (Porquet et al. 2004; Piconcelli et
al. 2004b; Schartel et al. 2004).

Alternatively, the change in the spectral properties between the
{\it ASCA} and the \xmm~observation could be explained in terms of a
temporal variation of the column density in the absorbing material  as
seen in some local AGNs (e.g. Akylas et al. 2002; Risaliti et
al. 2002; Lamer et al. 2003). This spectral behaviour suggests the
presence of clouds with different \nh~drifting across our line of
sight responsible of the observed \nh~variations.

%%%%%%%%%%%%% Estimated reflection <3
%%%%%%%%%%%%% with the same \chi2

\subsection{Fe K$\alpha$ emission line}
The Fe K$\alpha$ emission line detected at $\sim$ 3$\sigma$ in \rx~is
the most distant in a unlensed QSO observed so far (Chartas et
al. 2004). Even more interesting is the fact that it has been observed
in a Type II QSO, an elusive class of objects whose X--ray spectral
properties are not well known.  Bearing in mind the large errors
affecting both the measurements, the value of EW = 460$^{+330}_{-290}$
eV obtained by the \xmm~observation agrees  well with that reported
in RT00.  It appears, therefore, that the strength of the iron line
did not change between  the two observations.  This indicates that the
Fe emitting material traces the variability of the nuclear flux  and that the
fluorescence originates near the nucleus. Given the temporal
difference between the two observations we infer an upper limit of $<$
2.45 pc on the distance nucleus/Fe emitter.  The value of the EW is
unusually large if compared with the results of Page et al. (2004)
who reported a decrement of the mean EW with  increasing \lum. In
particular they predict an $\langle$EW$\rangle$\simlt~100 eV at  \lum
$\approx$10$^{45}$ \ergs~ i.e. the luminosity measured for \rx.  The
energy of the line indicates that it emerges from cold matter,
therefore an origin in the X--ray/optical absorbing material appears
likely. However, the best-fit value of the EW is too large to be
accounted by the amount of the absorption (few $\times$  10$^{21}$
\cm2) estimated for \rx. A plausible explanation for such an EW value
could be a  combination of different processes such as fluorescence in
transmission along the obscuring material plus reflection off the
inner walls of the optically thick torus and/or the accretion disk
into our line of sight. The detection of a continuum Compton-reflection
component in the \epic~spectrum is very marginal and we could
only obtain an upper limit of $R$(=$\Omega$/2$\pi$) $<$2 for the
covering factor of the material irradiated by the X-ray source, 
which is  still consistent with observed Fe K$\alpha$ line EW.
Since the latter  is consistent at 2$\sigma$ with a value  (EW $\sim$ 700
eV) typically observed in Compton--thick AGNs (Levenson et al. 2002;
Guainazzi et al. 2004) we also calculated the ratio $T$ = \fhx$/F_{OIII}$,
which is an indicator of the ``thickness'' of the absorbing matter
(Bassani et al. 1999) using the IR data in G99.  The resulting value
of $T \gg$ 1 rules out this hypothesis that the observed X--ray
continuum is entirely due to
reflection scattering.

In conclusion,  the X--ray spectral properties of \rx~obtained by this
\xmm~observation agree well with the predictions of the AGN Unified
Models concerning absorption. They also follow the trend of  X--ray
spectral slopes being constant with redshift or luminosity, as recently
inferred for large samples of RQ QSOs (e.g. P03; Vignali et al. 2003;
Piconcelli et al. 2004b).
%%%%%%%%%%%%%%%%%%%%%%%%%%%%%%%%%%%%%%%%%%%%%%%%%%%%%%%%%%%%%%%%%%%%%%%%%%%%%%%%%%%
\begin{acknowledgements}
We thank the anonymous referee for her/his interesting suggestions.
We gratefully acknowledge Marcus Kirsch and Pedro Rodriguez-Pascual for helpful comments. 
We would like to thank the staff of the \xmm~Science Operations Center for their support.
\end{acknowledgements}
%%%%%%%%%%%%%%%%%%%%%%%%%%%%%%%%%%%%%%%%%%%%%%%%%%%%%%%%%%%%%%%%%%%%%%%%%%%%%%%%%%%


\begin{thebibliography}{}
\bibitem {} Akylas, A., Geogakakis, A., \& Georgantopolous, I., 2004, MNRAS, in press
\bibitem {} Akiyama, M., Ohta, K., \& Yamada, T., et al., 2000, ApJ, 532, 700
\bibitem {} Almaini, O., Boyle, B. J., Griffiths, R. E., et al., 1995, MNRAS, 277, L31 (A95)
\bibitem {} Antonucci, R., 1993, ARA\&A, 31, 473
\bibitem {} Bassani, L., Dadina, M., Maiolino, R., et al., 1999, ApJS, 121, 473 
\bibitem {} Bennett, C.~L., Halpern, M., Hinshaw, G., et al., 2003, ApJS, 148, 1
\bibitem {} Cappi, M.,  Matsuoka, M., Comastri, A., et al., 1997, ApJ, 478, 492
\bibitem {} Chartas, G., Eracleous, M., Agol, E., \& Gallagher, S.~C., 2004, ApJ, 606, 78
\bibitem {} Dietrich, M.,  \& Hamann, F., 2004, ApJ, 611,  
\bibitem {} Fabian, A.~C., 2003, Proc. of "Carnegie Observatories Astrophysics Series, Vol. 1: Coevolution of Black Holes and Galaxies," ed. L.C. Ho (Cambridge Univ. Press)
\bibitem {} Fan, X., Hennawi, J. F., Richards, G.~T., et al., 2004, ApJ, in press(astro-ph/0405138)
\bibitem {} Ferrero, E., \& Brinkmann, W., 2003, A\&A, 402, 465
\bibitem {} Fiore, F., Elvis, M., Giommi, P., \& Padovani, P., 1998, ApJ, 429, 79
\bibitem {} Georgantopoulos, I., Almaini, O., Shanks, T., et al. 1999, MNRAS, 305, 125 (G99)
\bibitem {} Georgantopoulos, I., Georgakakis, A., Stewart, G. C., et al., 2003, MNRAS, 342, 321
\bibitem {} Georgantopoulos, I., Georgakakis, A., Akylas, A., et al., 2004, MNRAS, 352, 91
\bibitem {} Gilli, R., Salvati, M., \& Hasinger, G., 2001, A\&A 366, 407
\bibitem {} Granato, L., Danese, L., \& Franceschini, A., 1997, ApJ, 486, 147
\bibitem {} Grupe, D., Mathur, S., Wilkes, B., \& Elvis, M., 2004, ApJ, 127, 1
\bibitem {} Guainazzi, M., Fabian, A. C., Iwasawa, K., Matt, G., \& Fiore, F., 2004, MNRAS, (astro-ph/0409689)
\bibitem {} Jansen, F., Lumb, D., Altieri, B., et al., 2001, A\&A, 365, L1 
\bibitem {} Komossa, S., \& Fink, H., 1997, A\&A, 327, 555
\bibitem {} Lamer, G., Uttley, P., \& McHardy, I. M.,  2003, MNRAS 342, L41
\bibitem {} Levenson, N. A., Krolik, J. H., Zycki, P. T., et al., 2002, ApJ 573, 81
\bibitem {} Makishima, K., 1986, Lecture Notes in Physics, 266, 246
\bibitem {} Norman, C., Hasinger, G., Giacconi R., et al., 2002, ApJ, 571, 218
\bibitem {} Padovani, P., Allen, M.~G., Rosati, P., \& Walton, A., 2004, A\&A, in press (astro-ph/0406056)
\bibitem {} Page, K. L., O'Brien, P. T., Reeves, J. N., \& Turner, M. J., 2004, MNRAS, 347, 316
\bibitem {} Piconcelli, E., Cappi, M., Bassani, L., Di~Cocco, G., Dadina, M., 2003, A\&A, 412, 689 (P03)
\bibitem {} Piconcelli, E., Jimenez-Bailon, E., Guainazzi, M., et al., 2004a, MNRAS, 351, 161
\bibitem {} Piconcelli, E., Jimenez-Bailon, E., Guainazzi, M., et al., 2004b, A\&A in press, (astro--ph/0411051)
\bibitem {} Predehl, P., \& Schmitt, J.~H.~M.~M., 1995, A\&A   293, 889
\bibitem {} Porquet, D., Reeves, J.N., O'Brien, P., \& Brinkmann, W., 2004, A\&A, 422, 85
\bibitem {} Reeves, J.~N., Turner, M.~J.~L., 2000, MNRAS, 316, 234 (RT00)
\bibitem {} Schartel, N., Rodriguez-Pascual, P. M., Santos-Lleo, M., et al., 2004, A\&A,  in press
\bibitem {} Struder, L., Briel, U., Dennerl, K., et al., 2001, A\&A 365, L18
\bibitem {} Turner, M.~J.~L., Abbey, A., Arnaud, M., et al., 2001, A\&A, 365, L27
\bibitem {} Vignali, C., Comastri, A., Cappi, M., et al., 1999, ApJ, 516, 582
\bibitem {} Vignali, C., Brandt, W. N., Schneider, D. P., et al., 2003, AJ, 125, 2876
\bibitem {} Williams, O.R., Turner, M. J. L., Stewart, G. C., et al.,  1992, ApJ, 389, 157
\bibitem {} Zamorani, G., Henry, J.P., Maccacaro, T., et al., 1981, ApJ, 245, 357

\end{thebibliography}
\end{document}